\documentclass[prl,twocolumn,superscriptaddress,showpacs,amsmath,amssymb]{revtex4-2}
\usepackage{graphicx, color}
\usepackage{dcolumn}
\usepackage{bm}
\usepackage{epstopdf}
\usepackage{physics}
\usepackage[usenames,dvipsnames]{xcolor}
\usepackage{subfigure}
\usepackage{pifont}

\usepackage{verbatim}

\begin{document}
\title{Coupling the Higgs mode and ferromagnetic resonance in spin-split superconductors with Rashba spin-orbit coupling} 

\author{Yao Lu}
\affiliation{Department of Physics and Nanoscience Center, University of Jyv\"askyl\"a, P.O. Box 35 (YFL), FI-40014 University of Jyv\"askyl\"a, Finland}

\author{Risto Ojaj\"arvi}
\affiliation{Department of Physics and Nanoscience Center, University of Jyv\"askyl\"a, P.O. Box 35 (YFL), FI-40014 University of Jyv\"askyl\"a, Finland}

\author{P. Virtanen}
\affiliation{Department of Physics and Nanoscience Center, University of Jyv\"askyl\"a, P.O. Box 35 (YFL), FI-40014 University of Jyv\"askyl\"a, Finland}

 \author{M.A.~Silaev}
\affiliation{Department of Physics and Nanoscience Center, University of Jyv\"askyl\"a, P.O. Box 35 (YFL), FI-40014 University of Jyv\"askyl\"a, Finland}
\affiliation{Moscow Institute of Physics and Technology, Dolgoprudny, 141700 Russia}
\affiliation{Institute for Physics of Microstructures, Russian Academy of Sciences, 603950 Nizhny Novgorod, GSP-105, Russia}

\author{Tero T. Heikkil\"a}
\affiliation{Department of Physics and Nanoscience Center, University of Jyv\"askyl\"a, P.O. Box 35 (YFL), FI-40014 University of Jyv\"askyl\"a, Finland}

\date{\today}
\begin{abstract}
We show that the Higgs mode of superconductors can couple with spin dynamics in the presence of a static spin-splitting field and Rashba spin-orbit coupling.
The Higgs-spin coupling dramatically modifies the spin susceptibility near the superconducting critical temperature and consequently enhances the spin pumping effect in a ferromagnetic insulator/superconductor bilayer system. We show that this effect can be detected by measuring the magnon transmission rate and the magnon-induced voltage generated by the inverse spin Hall effect. 

\end{abstract}

\maketitle

Superconductors (SC) with broken $U(1)$ symmetry host two kinds of collective modes associated with the order parameter fluctuations: the phase mode and the amplitude mode. Coupled to a dynamical gauge field, the phase mode is lifted up to the plasma frequency \cite{phasemode1} due to the Anderson--Higgs mechanism \cite{anderson,higgs}. The other collective mode in SC is the amplitude mode \cite{Higgsmode1,particleSC} with an energy gap of 2$\Delta$, called the Higgs mode by analogy with the Higgs boson \cite{higgs} in particle physics. It was commonly believed that unlike the phase mode the Higgs mode usually does not couple linearly to any experimental probe. That is why in earlier experiments, the Higgs mode was only observed in charge-density-wave (CDW) coexisting systems \cite{littlewood1982amplitude,littlewood1981gauge,higgscdw1,higgscdw2,higgscdw3,higgscdw4}.  With the advance of terahertz spectroscopy technique \cite{Thz2} it became possible to investigate the Higgs mode through the nonlinear light--Higgs coupling \cite{Thzhiggs1,Thzhiggs2,Thzhigg3,mikhail1,mikhail2}.  In these experiments, the perturbation of the order parameter is proportional to the square of the external electromagnetic field $\text{\ensuremath{\delta\Delta\propto E^{2}}}$, so very strong laser pulses are required.

Recently, it has been shown that in the presence of a supercurrent the Higgs resonance can actually contribute to the total admittance $Y_{\Omega}$ due to the linear coupling of the Higgs mode and the external electromagnetic field \cite{moor2017amplitude,supercurrent1,supercurrent2,supercurrent3,supercurrent4}. This can be understood from a symmetry argument. Suppose the external electric field is linearly polarized in the $x$ direction $E=\hat{x}E_{x}e^{i\Omega t}$. The linear coupling of the Higgs mode and the external field is represented by the susceptibility $\chi_{\Delta E}=-\frac{\partial^{2}S}{\partial\Delta\partial E}$ obtained from the action $S$ describing the electron system containing the pair potential field $\Delta$. Without a supercurrent, the system preserves the inversion symmetry ($\hat{I}$) and the mirror symmetry in the $x$ direction ($\hat{M}_x$). On the other hand $\chi_{\Delta E}$ is odd under both these operations because $E$ changes sign under $\hat{I}$ and $\hat{M}_x$ whereas $\Delta$ remains the same. Therefore $\chi_{\Delta E}$ has to vanish. In the presence of a supercurrent, the inversion symmetry and the mirror symmetry are both broken and there is no restriction for $\chi_{\Delta E_{x}}$ from these symmetries, so $\chi_{\Delta E}$ can be nonzero. This symmetry argument also explains why the Higgs mode does not couple with an external field in the direction perpendicular to the supercurrent.

 \begin{figure}
\centering
\subfigure{\label{a}\includegraphics[width=\columnwidth]{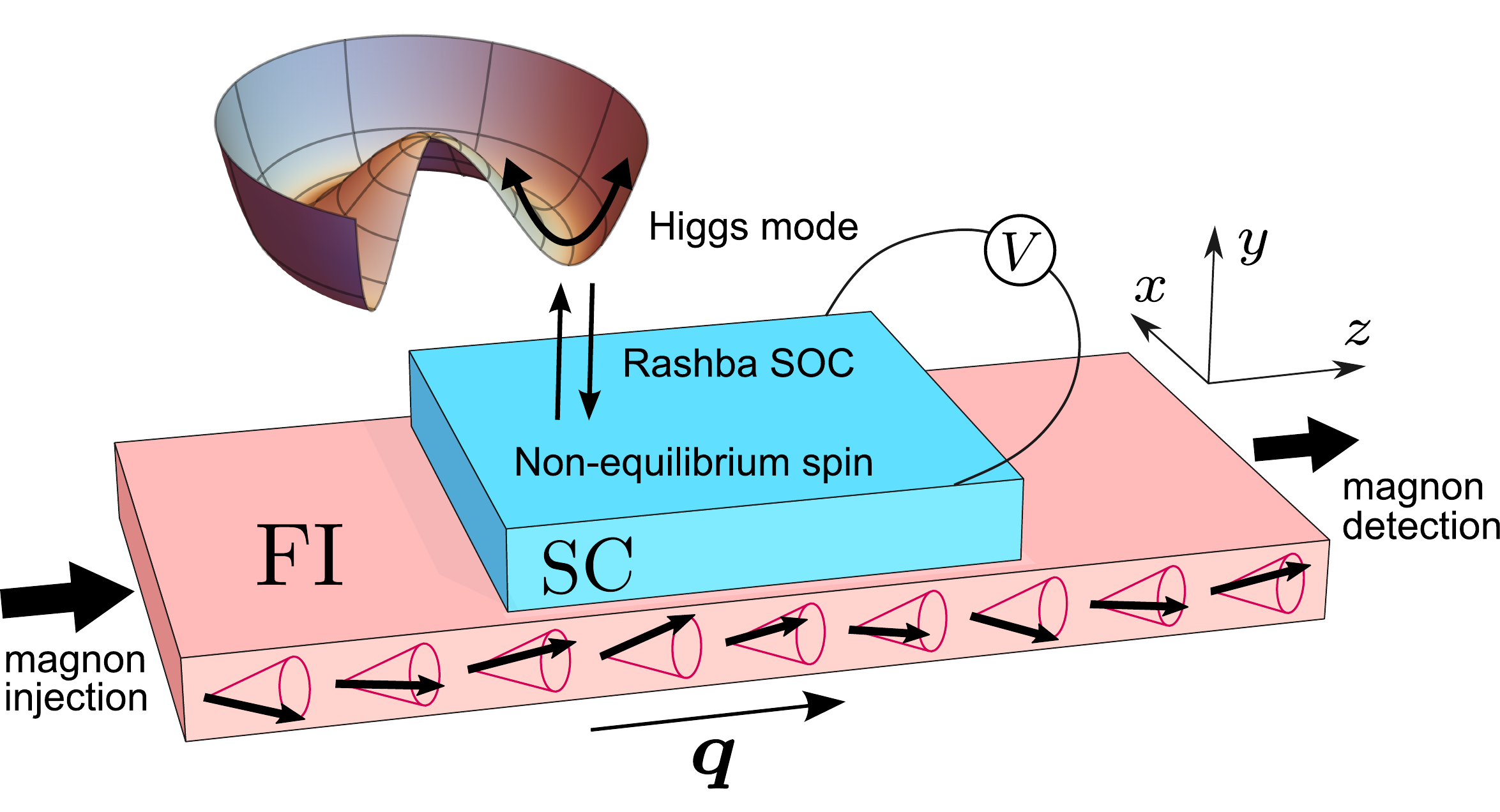}}
\caption{System under consideration. A superconductor thin film is placed on the top of a FI with in-plane magnetization. The SC and FI are coupled via spin exchange interaction. The magnon in FI can be injected into SC in a process known as the spin pumping effect. For magnon frequency $\Omega=2\Delta_0$ the SC Higgs mode greatly increases the spin pumping.}\label{Fig:setup}
\end{figure}

Now a natural question arises: without a supercurrent does the Higgs mode couple linearly with other external probes, such as spin exchange fields? As we show in this Letter it does. The above discussion indicates that the decoupling of the Higgs mode is protected by certain symmetries. In order to couple the Higgs mode to an external field one needs to break these symmetries. 
Here we show how it happens in a ferromagnetic insulator (FI)/superconductor (SC) bilayer system (Fig.~\ref{Fig:setup}). Magnons with momentum $\boldsymbol{q}$ and frequency $\Omega$ in the FI can be injected into the SC in a process known as spin pumping \cite{pumpingrmp1,pumpingnphys1,pumping1,pumping2,pumping3,pumping4,pumping5}. We predict that the Higgs mode in the SC couples linearly with the magnon mode in the FI in the presence of Rashba spin-orbit coupling and the magnetic proximity effect into the SC. In this system the symmetries protecting Higgs-spin decoupling are broken: in particular, the (spin) rotation symmetry and the time-reversal symmetry. Near the critical temperature, superconductivity is suppressed and $\Delta_0$ becomes comparable with the magnon frequency $\Omega$. When the magnon frequency matches the Higgs frequency $\Omega_{M}=2\Delta_0$, the Higgs mode is activated and the magnon absorption is hugely enhanced which can be detected through the inverse spin Hall effect (iSHE) \cite{iSHE1,iSHE2,iSHE3}. 
This effect can possibly explain the voltage peak observed in the experiment \cite{spinSebeck}.

We consider a SC/FI bilayer  in which the FI and the SC are coupled via the exchange interaction as shown in Fig.~\ref{Fig:setup}. For simplicity, we assume that the thickness $d$ of the SC film is much smaller than the spin relaxation length and the coherence length so that we consider it as a 2D system. The magnetization of the FI can be written as $\boldsymbol{m}=\boldsymbol{m}_0+\boldsymbol{m}_{\Omega}$, where $\boldsymbol{m}_0$ is the static manetization polarized in the $z$ direction and $\boldsymbol{m}_{\Omega}$ is the dynamical component perpendicular to $\boldsymbol{m}_0$. When magnons (spin waves) are excited in the FI, they can be injected into the SC in a process known as the spin pumping effect. The DC interface spin current flowing from the FI into the SC is polarized in the $z$ direction and given by \cite{ojajarvi2021}
\begin{equation}
    I_{z}=\sum_{\Omega,\boldsymbol{q}}-2J_{sd}\text{Im}[\tilde{\chi}_{ss}(\Omega,\boldsymbol{q})]m_{\Omega,\boldsymbol{q}}^2,
\end{equation}
where $J_{sd}$ is the exchange coupling strength and  $m_{\Omega}$ is the Fourier amplitude of $\boldsymbol{m}_{\Omega}$. $\tilde{\chi}_{ss}(\Omega,\boldsymbol{q})$ is the total dynamical spin susceptibility $\tilde{\chi}_{ss}(\Omega,\boldsymbol{q})=S_+(\Omega,\boldsymbol{q})/h_+(\Omega,\boldsymbol{q})$, where $\boldsymbol{S}$ is the dynamical spin of the SC, $\boldsymbol{h}$ is the proximity induced exchange field $\boldsymbol{h}=J_{sd}\boldsymbol{m}/d$ \cite{heikkila2019thermal} and for a vector ${\bm A}=(A_x,A_y,A_z)$ the $\pm$ component is defined as $A_{\pm}=A_x\pm iA_y$.
 One can see that for a fixed $J_{sd}$, the efficiency of the magnon injection is soley determined by $\tilde{\chi}_{ss}(\Omega,\boldsymbol{q})$. The spin susceptibility of superconductors has been extensively studied \cite{pumping5,silaev2020finite}. However the previous theories, based on the static mean-field description, 
failed to explain the peak of the iSHE signal observed in the spin Seebeck experiment \cite{spinSebeck}. In this work, we start with the general partition function of the SC, $Z=\int D[\bar{\Psi},\Psi,\bar{\Delta},\Delta]e^{-S}$ obtained by performing the Hubbard-Stratonovich transformation. The action $S$ is given by

\begin{eqnarray}
    S=\beta\sum_{K,Q}\bar{\Psi}_{K}\left(-i\omega+\epsilon_{\boldsymbol{k}}-\boldsymbol{h}\cdot\boldsymbol{\sigma}\right)\Psi_{K}+\Delta_{Q}\Psi_{K+Q}\Psi_{-K}\nonumber\\
    +\bar{\Delta}_{-Q}\bar{\Psi}_{K}\bar{\Psi}_{-K-Q}+\frac{\bar{\Delta}_{-Q}\Delta_{Q}}{U},
\end{eqnarray}
 Here $K=(\omega,\boldsymbol{k})$ and $Q=(\Omega,\boldsymbol{q})$ are the four-momenta of the electrons and magnons, 
 respectively. 
 $\omega=(2n+1)\pi T$ and $\Omega=2n\pi T$ are the Matsubara frequencies with $n\in Z$ and $\beta=1/T$. $\epsilon_{\boldsymbol{k}}$ is the energy dispersion of the electron in the normal state, $\boldsymbol{h}$ is the proximity induced exchange field, and $U$ is the BCS interaction. In the mean-field theory, one can ignore the path integral over  $\Delta$ and replace it by its saddle point value $\Delta_0$ which is determined by the minimization of the action $\frac{\partial S}{\partial\Delta}|_{\Delta=\Delta_0}=0$ after integrating out the fermion fields. 

To include the Higgs mode, we go beyond the mean-field theory and write the order parameter as $\Delta=\Delta_0+\eta$, where $\eta$ is the deviation of $\Delta$ from its saddle point value $\Delta_0$. Here we only consider the amplitude fluctuation of $\Delta$, so $\eta$ is real. Expanding the action to the second order in $\eta$ and the strength of the external Zeeman field $h_{\pm}$ gives $S=S_0-S_2$ with \cite{cea2015nonrelativistic}
\begin{equation}
   S_2=\beta\sum_{Q}\left[\begin{array}{cc}
\eta(-Q) & h_-(-Q)\end{array}\right]
\left[\begin{array}{cc}
-\chi_{\Delta\Delta}^{-1} & \chi_{\Delta s}\\
\chi_{s\Delta} & \chi_{ss}
\end{array}\right]\left[\begin{array}{c}
\eta(Q)\\
h_{+}(Q)
\end{array}\right].\label{eq:action_expansion}
\end{equation}
Here, all the susceptibilities are functions of $Q$. $S_0$ is the mean-field action without the external field. In usual superconductors the off-diagonal susceptibilities $\chi_{\Delta s}$ and $\chi_{s\Delta}$ vanish as required by the time-reversal symmetry and the (spin) rotation symmetry because these operations change the sign of $h_{+}$ but have 
no effect on $\eta$ \cite{supplementary,Mnote}. 
In the system under consideration, the proximity induced static exchange field breaks the time-reversal symmetry and RSOC breaks the (spin) rotation symmetry. Thus the pair-spin susceptibility does not 
have to vanish, allowing for a nonzero Higgs--spin coupling.

Then it is straightforward to calculate the total spin susceptibility $\tilde{\chi}_{ss}$ by integrating out the $\eta$ field
\begin{equation}
    \tilde{\chi}_{ss}=\chi_{ss}-\chi_{s\Delta}\chi_{\Delta\Delta}\chi_{\Delta s}.\label{eq:totspinsusc}
\end{equation}
The imaginary part of $\chi_{\Delta\Delta}$ is sharply peak at the Higgs frequency $\Omega=2\Delta$ dramatically modifying the total spin susceptibility.

{\em Phenomenological theory}. Before we go to the detailed calculations, we use a simple phenomenological theory to illustrate the effect of RSOC. It has been shown that RSOC can induce a  Dzyaloshinskii-Moriya (DM) interaction in superconductors described by the DM free energy \cite{silaev2021chiral}

\begin{equation}
F_{DM}=\sum_i\int d\boldsymbol{r}|\Delta|^2\boldsymbol{d}_{\alpha,i}\cdot(\boldsymbol{h}\times\nabla_i\boldsymbol{h}),
\end{equation}
 where both $\Delta=\Delta(\boldsymbol{r})$ and $\boldsymbol{h}=\boldsymbol{h}(\boldsymbol{r})$ are position dependent. $\boldsymbol{d}_{\alpha,i}$ is the DM vector proportional to the strength of spin-orbit coupling $\alpha$. For RSOC $\boldsymbol{d}_{\alpha}\propto\alpha[\sigma_x,-\sigma_z]$, where $\alpha$ is the spin-orbit coupling strength and $\sigma$ is the Pauli matrix acting on the spin space. To find the pair spin susceptibility we write $\Delta=\Delta_0+\eta(t)$, $\boldsymbol{h}=h_0\hat{z}+h_{+}(t)(\hat{x}+i\hat{y})$, where $\hat{
n}$ is the unit vector in the $n$ direction with $n=x,y,z$, and generalize the DM free energy to the time dependent DM action. Here we consider the case where the spin wave is propagating in the $z$ direction $h_+(t,\boldsymbol{r})=\sum_{\Omega,q_z}h_+(\hat{x}+i\hat{y})e^{i(\Omega t-q_zz)}$. Focusing on the first order terms in $\eta(t)$ and $h_{+}(t)$ and Fourier transforming them to momentum and frequency space, the DM action can be written as
 
 \begin{eqnarray}
     S_{DM1}=\beta\sum_{\Omega,q_z}iq_z\Delta_0h_0h_+(\Omega,q_z)\eta(\Omega,q_z)\tilde{\boldsymbol{d}}_{\alpha,z}(\Omega,q_z)\nonumber\\
     \cdot \left(i\hat{x}-\hat{y}\right)
     ,
 \end{eqnarray}
where $\tilde{\boldsymbol{d}}_{\alpha,i}$ is the dynamical DM vector, which  has the same finiteness and spin structure as $\boldsymbol{d}_{\alpha,i}$ from symmetry analysis. From the above expression, one can see that the Higgs mode couples linearly with the spin degree of freedom in the presence of RSOC. 

{\em Spin susceptibility}. We adopt the quasiclassical approximation to systematically evaluate the susceptibilities. In the diffusive limit, this system can be described by the Usadel equation \cite{moor2017amplitude,silaev2020finite, usadel1970generalized,usadel2,bergeret2005odd,SOC1,SOC2}
 \begin{figure}[h!]
\centering
\subfigure{\includegraphics[width = 1\columnwidth]{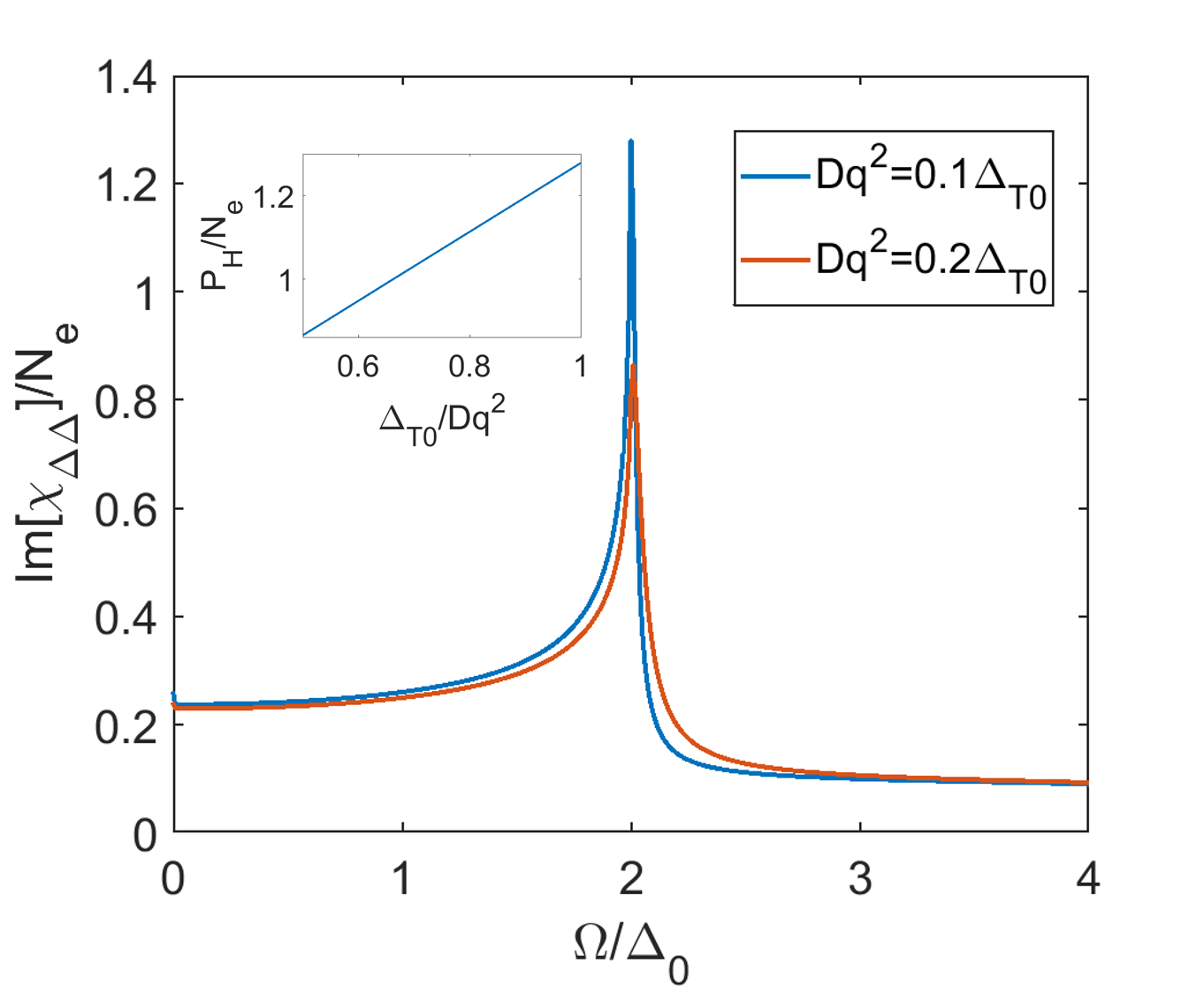}}
\caption{Imaginary part of  the pair susceptibility. This can be interpreted as the spectral weight of the Higgs mode. A significant peak emerges when the driving frequency matches the Higgs frequency $\Omega=2\Delta_0$. The inset shows the height of the Higgs peak $P_H$ as a function of the inverse of the momentum $\boldsymbol{q}$. Parameters:  $\Delta_0=0.8\Delta_{T0}$, $h_0=0.5\Delta_{T0}$ with $\Delta_{T0}\equiv\Delta_0(T=0)$. }\label{Fig:Higgs}
\end{figure}
\begin{equation}
-i\left\{ \tau_{3}\partial_{t},\hat{g}\right\} =D\tilde{\nabla}\left(\hat{g}\tilde{\nabla}\hat{g}\right)-i\left[H_0,\hat{g}\right]+\left[Xe^{i(\Omega t-q_zz)},\hat{g}\right].
\label{eq:Usadel}
\end{equation}
Here $\hat{g}$ is the quasiclassical Green function, $D=v_F\tau^2/3$ is the diffusion constant and $\tau$ is the disorder scattering time. $H_0=-ih_0\sigma_3+\Delta_0\tau_1$, where $h_0$ is the proximity induced effective static exchange field and $\tau_i$ is the Pauli matrix acting on the particle-hole space. $\tilde{\nabla}=(\tilde{\nabla}_z,\tilde{\nabla}_x)$ is the covariant derivative defined by $\tilde{\nabla}_z\cdot=\nabla_z+i\alpha[\sigma_x,\cdot]$, $\tilde{\nabla}_x\cdot=\nabla_x-i\alpha[\sigma_z,\cdot]$. The Usadel equation is supplemented by the normalization condition $\hat{g}^2=1$. In the quasiclassical approximation the approximate PH symmetry of the full Hamiltonian becomes exact. In the linear response theory, the external oscillating field $X$ is small and can be treated as a perturbation. Thus we can write the quasiclassical Green function as $\hat{g}=\hat{g}_{0}e^{i\omega(t_{1}-t_{2})}+\hat{g}_{X}e^{i(\omega+\Omega)t_{1}-i\omega t_{2}-iq_zz}$, where $\hat{g}_{0}$ is the static Green function and $\hat{g}_{X}$ is the perturbation of the Green function describing the response to the external field. Solving the Usadel equation we obtain the quasiclassical Green function, the anomalous Green function $F=N_e\text{Tr}\left[\tau_{1}\hat{g}\right]/4i$ and the $\sigma_+$ component of spin in the SC $\langle s\rangle=N_e\text{Tr}\left[\sigma_{-}\tau_{3}\hat{g}\right]/4i$, where $N_e$ is the electron density of states at the Fermi surface and $\text{Tr}$ is the trace. The susceptibilities can be evaluated as
\begin{equation}
\hat{\chi}=\left[\begin{array}{cc}
\chi_{\Delta\Delta}^{-1} & \chi_{\Delta s}\\
\chi_{s\Delta} & \chi_{ss}
\end{array}\right]=\left[\begin{array}{cc}
\frac{\partial F}{\partial\eta}+\frac{1}{U} & \frac{\partial F}{\partial h_+}\\
\frac{\partial\langle s\rangle}{\partial\eta} & \frac{\partial\langle s\rangle}{\partial h_+}
\end{array}\right].
\end{equation}

Let us first set $X=\Delta'\tau_1$ and consider the pair susceptibility. We assume the RSOC is weak and treat $\alpha$ as a perturbation. At $q=0$ and 0th order in $\alpha$, we have
\begin{equation}
      \chi_{\Delta\Delta}(i\Omega)=\left[\frac{N_eT}{2} \sum_{\omega,\sigma}\frac{4\Delta^2+\Omega^2}{s_{\sigma}(\omega)(4\omega^2-\Omega^2)}\right]^{-1},\label{eq:pairsusc1}
\end{equation}
where $s_{\sigma}(\omega)=\sqrt{(\omega+i\sigma h)^2+\Delta^2}$, with $\sigma=\pm 1$. To get the pair susceptibility as a function of real frequency, we need to perform an analytical continuation \cite{supplementary}. Thus $i\Omega$ is replaced by $\Omega+i0^+$. One can see that the 
$\chi_{\Delta\Delta}$ is peaked at the Higgs frequency $\Omega=2\Delta$.

We numerically calculate $\chi_{\Delta\Delta}$ with finite momentum and show the results in Fig.~\ref{Fig:Higgs} \cite{supplementary,code}. One can see that the imaginary part of the inverse of the pair susceptibility exhibits a sharp peak when the driving frequency equals $2\Delta_0$.  With a finite momentum, the Higgs mode is damped in the sense that the peak in the Higgs spectrum has a finite height and width.

\begin{figure}[h!]
\centering
\includegraphics[width=\columnwidth]{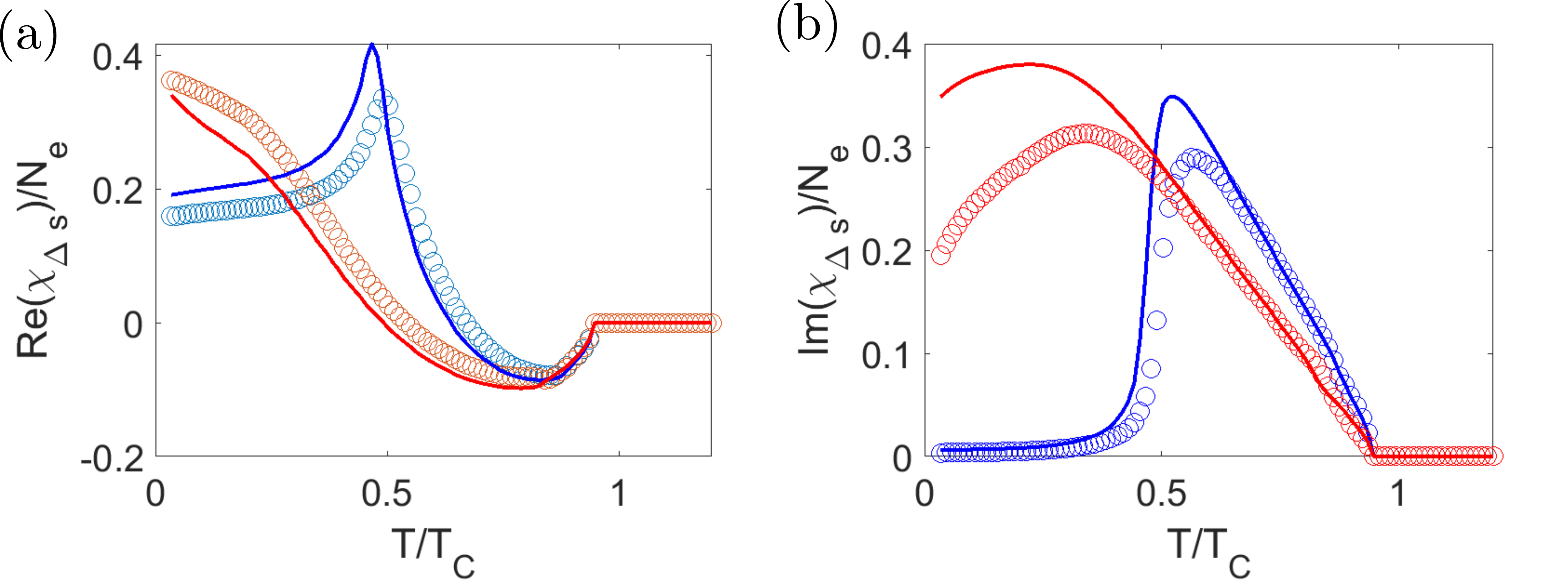}
\caption{Real part (a) and imaginary part (b) of pair-spin susceptibility. The solid line is the approximate result calculated from Eq.~(\ref{eq:pairspinsusc}) and the circles show the numerical solution from Eq.~(\ref{eq:Usadel}). Parameters used here are: $\Omega=0.8\Delta_{T0}$ for the blue lines, $\Omega=\Delta_{T0}$ for the red lines, $h_0=0.5\Delta_{T0}$, $Dq_z^2=D\alpha^2=0.01\Delta_{T0}$.
}\label{Fig:pairspinsusc}
\end{figure}

To study the response of this system to the external exchange field we set $X=h_{+}\sigma_+\tau_3$. Again we treat $\alpha$ as a perturbation and write the Green function as 
\begin{equation}
   \hat{g}=\hat{g}_{0}e^{i\omega(t_{1}-t_{2})}+(\hat{g}_{h0}+\hat{g}_{h\alpha})e^{i(\omega+\Omega)t_{1}-i\omega t_{2}-iq_zz},
\end{equation}
where $\hat{g}_{h0}$ is 0th order in $\alpha$ and $\hat{g}_{h\alpha}$ is first order in $\alpha$. The 0th order solution in $\alpha$ is given by \cite{supplementary}
\begin{equation}
    \hat g_{h0}=\hat{g}_{h00}\otimes\sigma_+=\frac{i [ \tau_3-\hat g_\uparrow(1) \tau_3 
  \hat g_\downarrow(2)] h_{\Omega}\sigma_+}{s_\uparrow(1)+s_\downarrow(2)},
  \end{equation}
where $\hat{g}_{\uparrow/\downarrow}=\frac{(\omega\pm ih_0)\tau_3+\Delta\tau_1}{s_{\uparrow/\downarrow}}$ and $s_{\uparrow/\downarrow}=\sqrt{(\omega\pm ih_0)^2+\Delta^2}$. $\hat{g}_{h00}$ is a $2\times 2$ matrix in the particle-hole space. Without doing detailed calculations, one can immediately see that $\chi_{\Delta s}$ has to vanish without RSOC because $\hat{g}_{h}$ has no $\sigma_0$ component. In this case the external exchange field cannot activate the Higgs mode. To get a finite pair-spin susceptibility we need to consider the first order terms in $\alpha$ which break the spin rotation symmetry. The first order solution in $\alpha$ yields $\hat{g}_{h\alpha}={\rm diag}(\hat g_{h\alpha\uparrow}, \hat g_{h\alpha\downarrow})$ with 
\begin{equation}
    \hat{g}_{h\alpha\uparrow/\downarrow}=2iD\alpha\frac{\hat{g}_{0\uparrow/\downarrow}\left[\hat{g}_{h00},\hat{g}_{0\uparrow/\downarrow}\right]}{s_{\uparrow/\downarrow}(\omega_{1})+s_{\uparrow/\downarrow}(\omega_{2})}. \label{eq:pairspinsusc}
\end{equation}

\begin{figure}[h!]
\centering
\includegraphics[width=0.96\columnwidth]{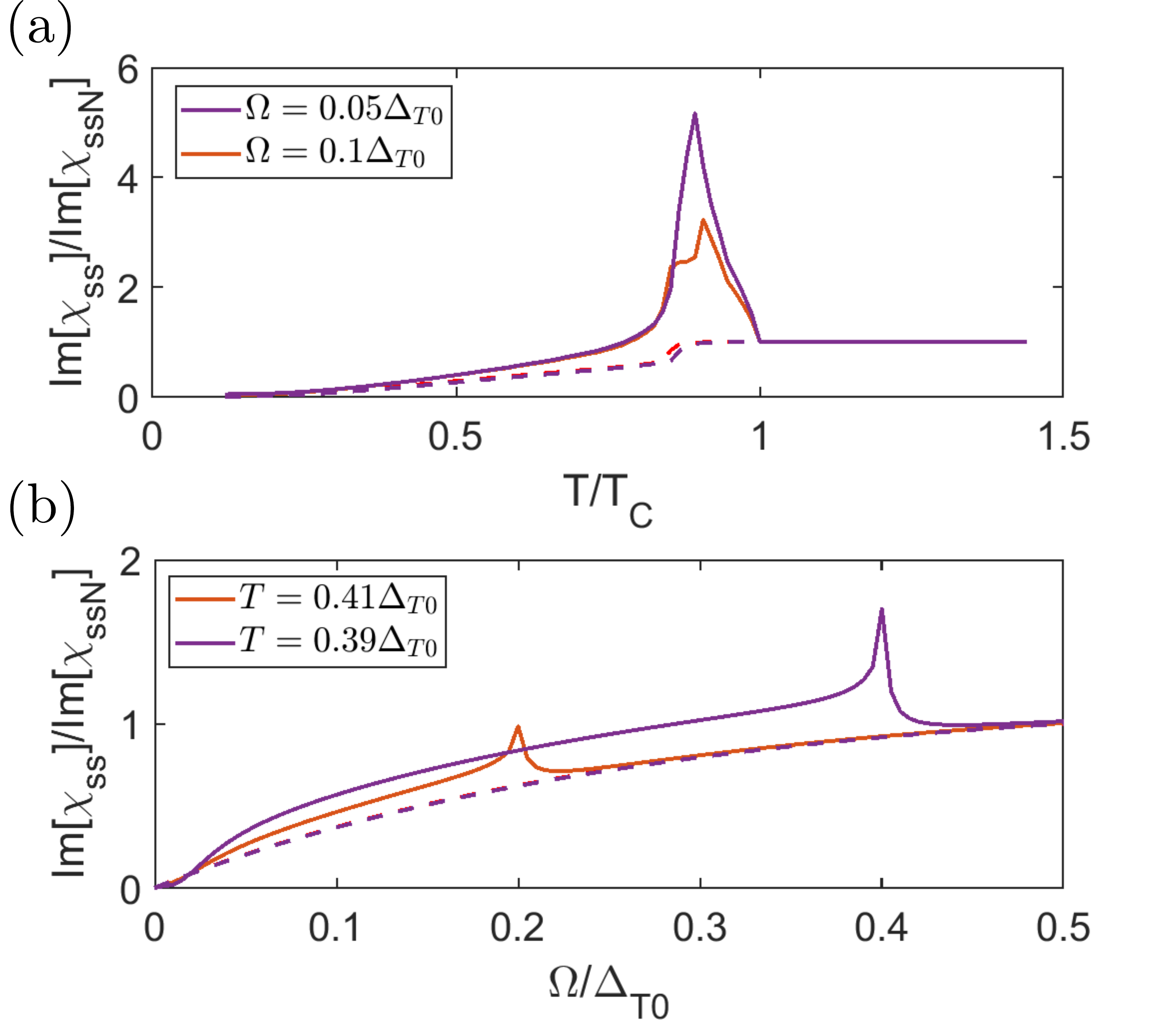}
\caption{(a) Total spin susceptibility as a function of temperature with a fixed frequency. (b) Total spin susceptibility as a function of frequency with a fixed temperature. The two temperatures have been chosen so that $\Delta(T_1)=0.2\Delta_{T0}$ and $\Delta(T_2)=0.1 \Delta_{T0}$.
The Higgs peak thus shows up when $\Omega=2 \Delta(T)$. The parameters used here are: $h_0=0.5\Delta_{T0}$, $Dq_z^2=D\alpha^2=0.01\Delta_{T0}$.}\label{Fig:totchis}
\end{figure}


Since the 0th order term does not contribute to the pair-spin susceptibility, we have $\chi_{\Delta s}=\text{Tr}[\tau_1\hat{g}_{h\alpha}]/4ih_{+}$. We compare this analytical result with the non-perturbative numerical solution of the Usadel equation in Fig.~\ref{Fig:pairspinsusc}. It shows that the perturbative approach is accurate at high temperatures when $D\alpha^2\ll\Delta_0,T$, and captures the qualitative behavior of $\chi_{\Delta s}$ also at the low temperatures. Another feature of this pair spin susceptibility is that at a lower frequency ($\Omega=0.8\Delta_{T0}$), $\chi_{\Delta s}$ is suppressed at low temperatures because the spin excitation is frozen by the pair gap at low temperatures. On the other hand, at higher frequency ($\Omega=\Delta_{T0}$), $\chi_{\Delta s}$ is slightly enhanced at low temperatures. 

We can also get the bare spin susceptibility from $\hat{g}_{h0}$, $\chi_{ss}=\text{Tr}[\sigma_-\tau_3\hat{g}_{h0}]/4ih_{+}$. Then it is straightforward to calculate the total spin susceptibility according to Eq.~(\ref{eq:totspinsusc}). The results are shown in Fig.~\ref{Fig:totchis}. The total spin susceptibility exhibits a significant peak near critical temperature. This is a signature of the Higgs mode with the frequency $\Omega=2\Delta_0$. The dependence of the total susceptibility on the strength of RSOC is studied in the supplementary information \cite{supplementary}. The details depend sensitively on the amount of disorder, as in the disordered case increasing RSOC leads to a stronger spin relaxation. We note that even though the pair-spin susceptibility is linear in momentum $q_z$, the magnon momentum need not be large for the detection of the Higgs mode. This is because the spectral weight of the Higgs mode is proportional to $1/q_z^2$ at the Higgs frequency, so that the height of the peak in the total spin susceptibility is independent of the magnon momentum. 

{\em Experimental detection}. We propose that the Higgs mode in Rashba superconductors can be detected in the spin pumping experiment as shown in Fig.~\ref{Fig:setup}. Magnons in the FI with momentum $\boldsymbol{q}$ and frequency $\Omega$ are  injected from one side of FI and propagate in the $z$ direction towards the other end. Due to the spin pumping effect, part of the magnons can be absorbed by the SC on top of it and converted to quasiparticles. This spin injection causes a spin current $I_s$ flowing in the out-of-plane direction. In the presence of RSOC, $I_s$ is 
converted into a charge current $I_e$ via the iSHE $I_e=\theta_{xz}^z I_s$, where $\theta$ is the spin Hall angle \cite{tokatly2017usadel}. When the width of the SC is smaller than the charge imbalance length the non-equilibrium charge accumulation cannot be totally relaxed resulting into a finite resistance $\rho$ of the SC. Therefore a voltage can be measured across the SC, given by

\begin{equation}
    V=\theta_{xz}^z\rho\sum_{\Omega,\boldsymbol{q}}-2J_{sd}\text{Im}[\tilde{\chi}_{ss}(\Omega,\boldsymbol{q})]m_{\Omega}^2.
\end{equation}
Thus by tuning the temperature or the frequency of magnon, one can observe a peak in the voltage \cite{spinSebeck}. Meanwhile we can also obtain the magnon absorption rate defined as the energy of the absorbed magnons divided by time

\begin{equation}
    W=2\Omega\sum_{\Omega,\boldsymbol{q}}-2J_{sd}\text{Im}[\tilde{\chi}_{ss}(\Omega,\boldsymbol{q})]m_{\Omega}^2.
\end{equation}
This magnon absorption rate results in a dip in the magnon transmission rate which is experimentally measurable.

{\em Conclusion}. In this Letter, we consider a FI/SC bilayer with RSOC in the bulk of the SC. Using symmetry arguments and microscopic theory, we show that the Higgs mode in the SC couples linearly with an external exchange field. This Higgs--spin coupling hugely enhances the total spin susceptibility near a critical phase transition point, which can be detected using iSHE or via strong frequency dependent changes in the magnon transmission. Note that in this work, we consider the diffusive limit where the disorder strength is stronger than the RSOC and exchange field. However, our conclusion on Higgs--spin coupling should still be valid in the case of strong RSOC. In fact, we expect that the coupling is much stronger with strong SOC in the clean limit. In the diffusive limit, the RSOC together with disorder effectively generate spin relaxation which reduces the proximity induced exchange field suppressing the Higgs--spin coupling. On the other hand, in the clean case without disorder this effect is absent and hence the Higgs--spin coupling can be stronger. We also compare the Higgs mode in the diffusive limit and ballistic limit in the supplemental material \cite{supplementary}.
\begin{acknowledgments}
This  work  was  supported  by Jenny and Antti Wihuri Foundation, and the  Academy  of  Finland  Project  317118 and the  European  Union’s  Horizon  2020  Research  and  Innovation Framework  Programme  under  Grant  No. 800923 (SUPERTED).
\end{acknowledgments}

\bibliography{refs}

\begin{thebibliography}{47}
\bibitem [{Mno()}]{Mnote}%
  \BibitemOpen
  \href@noop {} {}\bibinfo {note} {In general $h_+$ is inhomogeneous and the
  gradient of $h_+$ breaks the mirror symmetry locally. Thus we do not need to
  consider the mirror symmetry of the SC.}\BibitemShut {Stop}%
\end{thebibliography}

\end{document}